\documentclass[twocolumn,showpacs,preprintnumbers,amsmath,amssymb]{revtex4}
\usepackage{graphicx}
\usepackage{dcolumn}
\usepackage{bm}

\begin{document}


\title{Stationary solution of a weak-driven open Jaynes-Cummings system of a degenerate two-level atom coupled to an arbitrary-polarized cavity field}

\author{Sungsam Kang}
\author{Youngwoon Choi}
\author{Sooin Lim}
\author{Wookrae Kim}
\author{Jai-Hyung Lee}
\author{Kyungwon An}

\email{kwan@phya.snu.ac.kr}%
\affiliation{Department of Physics and Astronomy, Seoul National
University, Seoul, 151-742, Korea}

\date{\today}

\begin{abstract}
Analytical solution for the stationary density matrix is derived, by using the Morris-Shore transformation, for an open Jaynes-Cummings system of a two-level atom with Zeeman sublevel degeneracy coupled to an arbitrary-polarized cavity mode. In the limit of weak excitation with the number of quantum in the system not exceeding one, we have obtained the stationary solution of the master equation up to the first order of the driving field intensity. We have also derived the analytic expressions for the excitation spectra of atomic spontaneous emission and cavity transmission. Our results show that the system can be regarded as a non-degenerate two-level system with a single effective coupling constant which depends only on the elliptic angle of the driving field as long as the atom-cavity coupling is not too strong. A precise condition for this approximation is derived. This work provides a theoretical ground for experimentally realizing a Jaynes-Cummings system with a coupling constant continuously varied for various cavity quantum electrodynamics studies.
\end{abstract}


\maketitle

\section{\label{sec:sec1}INTRODUCTION}
In cavity quantum electrodynamics (QED), the Jaynes-Cummings model (JCM) \cite{ref:JCM_1st} is one of the key elements, describing the interaction between a two-state atom and a radiation mode of a cavity. It is not too much to say that the cavity QED has been developed within the theoretical framework of the JCM. In studying the JCM, one is often faced with the difficulty in dealing with multi-state atoms in actual problems \cite{ref:JCM_review}; the JCM itself is for a two-state single atom coupled to radiation modes. Researchers have proposed extended models for specific types of multi-energy-level structure such as three-level atoms \cite{ref:JCM_three_level_general, ref:JCM_three_damp}, ladder-type four-level atoms \cite{ref:JCM_ladder_4}, two-level atoms with Zeeman degeneracy \cite{ref:Kimble_multi, ref:degen_sphere}, etc.

Among those multi-state atoms, a two-level atom with Zeeman sublevel degeneracy, or the so-called a {\em degenerate} two-level system, is of particular interest owing to its possible application to quantum state engineering \cite{ref:appl_map, ref:appl_pump} and quantum information processing \cite{ref:appl_pspod, ref:appl_remp}. Unfortunately, finding  an analytic closed-form description of the interaction of the degenerate two-level system with a resonant cavity field is a very complicated matter because of the coherence between the Zeeman sublevels. That is why this system has usually been studied by numerical methods \cite{ref:Kimble_multi} except for some limited cases with specific polarization \cite{ref:appl_map, ref:appl_remp} or specific angular momentum \cite{ref:degen_sphere}.

Nonetheless, as speculated in the review article of Ref.\ \cite{ref:JCM_review}, it might be possible to find a stationary solution for the interaction of a multi-level atom with an arbitrary-polarized cavity field in an invariant and analytic form with the aid of a unitary transformation introduced by Morris and Shore \cite{ref:MS_Transform}. For the case of a degenerate two-level atom interacting with an arbitrarily polarized light field in a free space, the problem in the semi-classical limit, where the atom is quantum mechanical but the field is regarded classical, comes down to finding a steady-state solution for the generalized optical Bloch equation(GOBE) \cite{ref:GOBE}.
Researchers paid a particular attention to this problem in order to describe polarization gradient cooling \cite{ref:subDoppler} and coherent population trapping with elliptical dark states \cite{ref:milner_EDS}, and a full theoretical analysis was made \cite{ref:Yudin_1st, ref:milner_pra} based on the Morris-Shore transformation \cite{ref:MS_Transform}.

In this paper, we have extended the previous works based on GOBE to an open JCM in a full-quantized manner. We consider the Zeeman sublevels degenerate in the absence of external magnetic fields. The cavity is weakly driven by an arbitrary-polarized field while the cavity mode is strongly coupled to the $F\leftrightarrow F+1$ closed transition of atomic hyperfine structure. We have found an analytic expression for the stationary density matrix by solving the master equation for the dominant order of the driving field intensity. We have also derived the analytic expressions for the excitation spectra of atomic spontaneous emission and cavity transmission, respectively. Our results show that the system can be regarded as a {\em non-degenerate} two-level system, unless the atom-cavity coupling is not too strong, with a single effective coupling constant which depends only on the elliptic angle of the driving field. We present a precise condition for this approximation to be valid. By using our results one should be able to realize a JCM with a coupling constant continuously varied for various cavity-QED experiments.

This paper is organized as follows. We formulate our system of interest with the master equation in Sec.\ \ref{sec:sec2} and then present a stationary-state solution by introducing natural basis in Sec.\ \ref{sec:sec3}. The excitation spectra of the cavity transmission and the atomic spontaneous emission and the resulting non-degenerate two-level system are discussed in Sec.\ \ref{sec:sec4}, followed by a conclusion in Sec.\ \ref{sec:sec5}.

\section{THEORETICAL MODEL} \label{sec:sec2}
\begin{figure}
\includegraphics[width=3.4in]{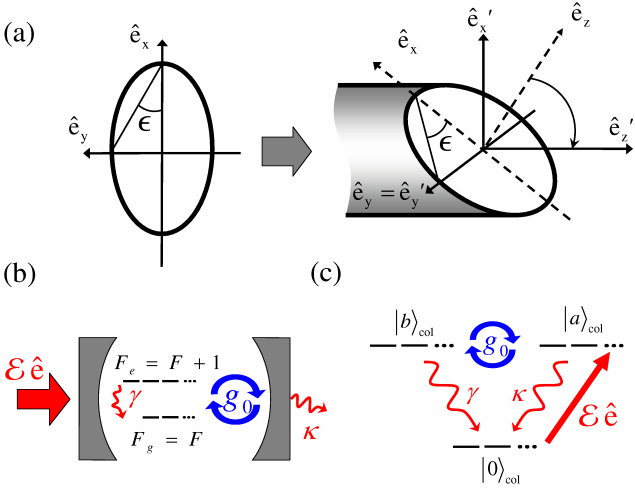}
\caption{(a) Left: conventional coordinate frame for elliptic polarization. Right: natural coordinate frame. Polarization vector is a superposition of circular and linear polarization. (b) Degenerate two-level atom interacting with a common cavity field. (c) Collective dressed-state representation with the number of quantum limited up to one.}
\label{fig:fig1}
\end{figure}

Arbitrary polarization in three dimensions can be expressed as
\begin{equation}
\hat{\mathbf{e}}=\sum_{q=-1}^{+1}e_{q}{\hat{\mathbf{e}}_{q}}
\end{equation}
where $\hat{\mathbf{e}}_q$'s are unit vectors. Note that in Cartesian coordinates, these unit vectors are chosen as
$\hat{\mathbf{e}}_0=\hat{\mathbf{e}}_z$, $\hat{\mathbf{e}}_\pm=\mp
(\hat{\mathbf{e}}_x \pm i \hat{\mathbf{e}}_y)/\sqrt{2}$.
It is always possible to remove one component of the polarization vector through an appropriate choice of $\hat{\mathbf{e}}_z$ axis. For example, if we choose $\hat{\mathbf{e}}_z$ axis to be perpendicular to the plane of polarization, the $\hat{\mathbf{e}}_z$ component vanishes, and the polarization becomes elliptical in general as shown in the left part of Fig.\ \ref{fig:fig1}(a).

Throughout this paper, we choose a {\em natural} coordinate frame as introduced in Ref.\ \cite{ref:Yudin_nc}. The unit vectors $\hat{\mathbf{e}}_i '$ of this coordinate frame are shown in the right part of Fig.\ \ref{fig:fig1}(a). In the natural coordinate frame, one of the circular component vanishes, and thus the polarization vector is a superposition of a circular and a linear polarization component:
\begin{equation}
\hat{\mathbf{e}}=\hat{\mathbf{e}}'_0 \sqrt{\cos 2\epsilon}
-\hat{\mathbf{e}}'_{+1}\sqrt{2} \sin \epsilon ,
\label{eq:el_pol}
\end{equation}
where $\epsilon$ is the elliptic angle defined by Fig.\ \ref{fig:fig1}(a), satisfying $-\pi/4 \leq \epsilon \leq +\pi/4$. In this frame, linear polarization can also be described naturally ($\epsilon=0$), and moreover, the analytic expressions of the matrix elements for the stationary density matrix become greatly simplified as to be shown below.

Our system of interest is shown in Fig.\ \ref{fig:fig1}(b). A degenerate two-level atom is coupled to a cavity with a representative coupling constant $g_0$ while the cavity is weakly driven by a classical field of arbitrary polarization $\hat{\mathbf{e}}$ and an amplitude $\cal{E}$. The decay rates of the cavity and the atom are denoted by $\kappa$ and $\gamma$ (both half widths), respectively. We assume that the frequency of the driving field ($\omega_L$) and that of the cavity ($\omega_C$) are both near resonant to the atomic transition ($\omega_A$), $(F_g =F)\leftrightarrow (F_e =F+1)$. The coupling constant $2g_0$ is the vacuum Rabi frequency for the cycling transition, $(m_F =F) \leftrightarrow (m_F'=F+1)$.

Under the rotating wave approximation, the Hamiltonian for this system can be described by a multi-level expansion of the Jaynes-Cummings Hamiltonian \cite{ref:Kimble_multi}. In the rotating frame of $\omega_L$ ({\em i.e.}, in the interaction picture), the Hamiltonian is written as
\begin{eqnarray}
H/\hbar&=&\Delta_C a^{\dag}a +\Delta_A \sum_{q=-1}^{+1}D_q^{\dag}D_q\nonumber\\
& &~ + g_0(V a^{\dag}+a V^{\dag})+{\cal{E}}(a+a^{\dag})\;,
\label{eq:Hamiltonian}
\end{eqnarray}
where $\Delta_{A, C}=(\omega_{A, C}-\omega_L)$ are the detunings of the atom and the cavity with respect to the driving field, respectively. The classical-driving-field amplitude $\cal{E}$ is scaled in such a way that ${\cal{E}}/\kappa$ represents a dimensionless injected photon flux. Operator $D_q$ is defined as an atomic lowering operator for $\hat{\mathbf{e}}_q '$ polarization
\begin{equation}
D_q =\sum_{m_F , m_F '}C^{F_e m_F '}_{F_g m_F 1 q} |g, m_F\rangle \langle e,m_F '|\;,
\label{Dq-defined}
\end{equation}
with the Clebsch-Gordan coefficients
\begin{equation}
C^{JM}_{j m j' m'}=(-1)^{j-j'+M}\sqrt{2J +1}
\left(
\begin{array}{ccc}
j&j'&J\\
m&m'&-M \end{array}
\right)\;.
\end{equation}
Note that $D_q$ satisfies the relation $\sum_{q}D_q ^{\dag}D_q=\sum_{m_F'}|e,m_F' \rangle\langle e, m_F'|$, which implies the conservation of the total population for a closed transition. Operator $V$ is an atomic lowering operator for $\hat{\mathbf{e}}$ polarization and can be written as
\begin{equation}
V=(\sum_{q=-1}^{+1} D_q \hat{\mathbf{e}}_q ')\cdot \hat{\mathbf{e}}
\label{V-defined}
\end{equation}
Since $m_F=-F, -F+1, \ldots, F$, and $m_F'=-F-1, -F, \ldots, F+1$, both operators $D_q$ and $V$ can be represented by $(2F+1) \times (2F+3)$ matrices. The time evolution of the system is then described by the master equation,
\begin{eqnarray}
\dot{\rho}&=&{\cal{L}} [\rho]=\frac{1}{i\hbar}[H, \rho]+\kappa(2a\rho a^{\dag}-a^{\dag}a\rho-\rho a^{\dag}a) \nonumber \\
& & +~\gamma \sum_{q=-1}^{+1}(2D_q \rho D_q^{\dag}-D_q^{\dag}D_q \rho-\rho D_q^{\dag}D_q ),
\label{eq:master}
\end{eqnarray}
and the stationary density matrix is obtained by solving ${\cal{L}}[\rho_{ss}]=0$.

\section{Weak Excitation Limit in Collective Dressed-State Basis}
In general, an analytic expression for the stationary solution of Eq.\ (\ref{eq:master}) cannot be obtained because the field operators $a$ and $a^{\dag}$ have an infinite number of bases. However, in the weak excitation limit, one can expect that the infinite number of field bases can be truncated to a finite number of low-quantum bases and that the steady-state solution can be expressed approximately in terms of a few low orders of driving-field amplitude $\cal{E}$ \cite{ref:Carmichael_ss_sol, ref:Brecha_ss_sol}. Under this assumption, the system can be described by the interaction between two manifolds of $(2F+1)$ collective dressed states $|a\rangle_{\textrm{col}}$ and $(2F+3)$ collective dressed states $|b\rangle_{\textrm{col}}$ with the number of quantum for both to be one. These manifolds are connected to a manifold of $(2F+1)$ collective ground dressed states $|\,0\,\rangle_{\textrm{col}}$ with no quantum through two decay channels as shown in Fig.\ \ref{fig:fig1}(c). These collective state vectors are defined as
\begin{subequations}
\label{eq:col_state}
\begin{eqnarray}
|{0}\rangle_{\textrm{col}}&=\sum_{m_F} \xi_{m_F}^{\,0}|g, m_F \rangle_{\rm atom} \otimes |0\rangle_{\rm field} \\
|{a}\rangle_{\textrm{col}}&=\sum_{m_F} \xi_{m_F}^{a}|g, m_F \rangle_{\rm atom} \otimes |1\rangle_{\rm field} \\
|{b}\rangle_{\textrm{col}}&=\,\sum_{m_F'} \xi_{m_F'}^{b}|e, m_F'\rangle_{\rm atom} \otimes |0\rangle_{\rm field}\;,
\end{eqnarray}
\end{subequations}
where \{$\xi_{i}^{(\,0, a, b)}$\} are complex amplitudes. There are as many as $(2F_{g,e}+1)$ independent sets of $\{\xi_{i}^{(\,0, a, b)}\}$. In the absence of the driving field (${\cal{E}}=0$), only \{$\xi_{i}^{\,0}$\} can have non-zero values
in the steady state. Therefore, we expect that the dominant terms of \{$\xi_i^{a,b}$\} are at least of the first order of $\cal{E}$ while \{$\xi_i^{\,0}$\} are of the zeroth order of $\cal{E}$.

With these bases the Hamiltonian and the density matrix can respectively be expressed by a $(6F+5)\times(6F+5)$ matrix, which can be divided into 9 sub-matrices, as
\begin{equation}
H=\hbar\left[\begin{array}{ccc}
\Delta_A {I}_{2F+3}&g_0 V^{\dag}&0\\
g_0 V&\Delta_C {I}_{2F+1}&{\cal{E}} I_{2F+1}\\
0&{\cal{E}} I_{2F+1}&0\end{array}\right],
\label{eq:Hamiltonian_M}
\end{equation}
\begin{equation}
\rho=\left[\begin{array}{ccc}
\rho_{bb}&\rho_{ba}&\rho_{b0}\\
\rho_{ab}&\rho_{aa}&\rho_{a0}\\
\rho_{0b}&\rho_{0a}&\rho_{00}\end{array}\right],
\label{eq:density}
\end{equation}
where ${I}_n$ denotes an $n\times n$ unit matrix and $\rho_{ij}=|{i}\rangle_{\textrm{col}}\langle{j}|_{\textrm{col}}$
for $\{i,j\}=\{\,0, a,b\}.$ By substituting Eqs.\ (\ref{eq:Hamiltonian_M}) and (\ref{eq:density}) into Eq.\ (\ref{eq:master}), we obtain a set of first-order differential equations
\begin{subequations}
\label{eq:GOBE_dia}
\begin{equation}
\dot{\rho}_{bb}=-2\gamma \rho_{bb}-ig_0 (V^{\dag}\rho_{ab}-\rho_{ba}V)\label{subeq_dia_bb}
\end{equation}
\begin{equation}
\dot{\rho}_{aa}=-2\kappa \rho_{aa}-ig_0 (V\rho_{ba}-\rho_{ab}V^{\dag})-i{\cal{E}}(\rho_{0a}-\rho_{a0})\label{subeq_dia_aa}
\end{equation}
\begin{equation}
\dot{\rho}_{00}=2\kappa\rho_{aa}+2\gamma\sum_{q}D_q \rho_{bb}D_q^{\dag}+i{\cal{E}}(\rho_{0a}-\rho_{a0})\label{subeq_dia_00}
\end{equation}
\end{subequations}
\begin{subequations}
\label{eq:GOBE_offdia}
\begin{equation}
\dot{\rho}_{ab}=(\dot{\rho}_{ba})^{\dag}=i(E_A^* -E_C)\rho_{ab}-ig_0(V\rho_{bb}-\rho_{aa}V)-i{\cal{E}}\rho_{0b}\label{subeq_ab}
\end{equation}
\begin{equation}
\dot{\rho}_{0b}=(\dot{\rho}_{b0})^{\dag}=iE_A^* \rho_{0b}+ig_0 \rho_{0a}V -i{\cal{E}}\rho_{ab} \label{subeq_0b}
\end{equation}
\begin{equation}
\dot{\rho}_{0a}=(\dot{\rho}_{a0})^{\dag}=iE_C^* \rho_{0a}+ig_0 \rho_{0b}V^{\dag}+i{\cal{E}}(\rho_{00}-\rho_{aa}),
\label{subeq_0a}
\end{equation}
\end{subequations}
where $E_A=\Delta_A-i\gamma$ and $E_C=\Delta_C-i\kappa$ are the complex energies of the atom and the cavity, respectively.

To solve for a stationary solution, we should find 9 sub-matrices which make the time derivatives in the left-hand side of Eqs.\ (\ref{eq:GOBE_dia}) and (\ref{eq:GOBE_offdia}) vanish, under the normalization condition $\mathrm{Tr}[\rho]=1$.
By rearranging terms and removing the higher order terms of $\cal{E}$ than the second order, we obtain
\begin{subequations}
\label{eq:off_dia_ss}
\begin{eqnarray}
\rho_{ab}=(\rho_{ba})^{\dag}&=&i s^* g_0 (V\rho_{bb}-\rho_{aa}V) \\ \nonumber
& &+\; is^* g_0 {\cal{E}}^2 \rho_{00}V{X}^{-1}
+O({\cal{E}}^4),
\label{subeq_ab_f}
\end{eqnarray}
\begin{equation}
\rho_{0b}=(\rho_{b0})^{\dag}=g_0
{\cal{E}}\rho_{00}V{X}^{-1}+O({\cal{E}}^3),
\label{subeq_0b_f}
\end{equation}
\begin{equation}
\rho_{0a}=(\rho_{a0})^{\dag}=-E_A^*
{\cal{E}}\rho_{00}{Y}^{-1}+O({\cal{E}}^3),
\label{subeq_0a_f}
\end{equation}
\end{subequations}
where $s=i/(E_A -E_C^*)$. $X$ and $Y$ are the square matrices
defined as
\begin{eqnarray}
{X}&=&E_A^* E_C^* ~{I}_{2F+3} -g_0^2 V^{\dag}V\;, \nonumber\\
{Y}&=&E_A^* E_C^* ~{I}_{2F+1} -g_0^2 VV^{\dag}\;.\nonumber
\end{eqnarray}
One can easily prove $X$ and $Y$ satisfy the following identities:
\begin{eqnarray}
&{X}V^{\dag}=V^{\dag}{Y},\;\;V{X}={Y}V.
\label{eq:VX_YV}
\end{eqnarray}

By substituting Eqs.\ (\ref{eq:off_dia_ss}) and (\ref{eq:VX_YV}) into Eq.\ (\ref{eq:GOBE_dia}), we then obtain\begin{subequations}
\label{eq:dia_ss}
\begin{eqnarray}
-2\gamma\rho_{bb}&-&\big[s^*g_0^2 V^{\dag} (\rho_{aa}-{\cal{E}}^2\rho_{00}Y^{-1}) V \nonumber\\
& & \hspace{0.5in} - s^*g_0^2V^{\dag}V\rho_{bb}+\textrm{h.c.} \big]=0,\hspace{0.2in}
\label{subeq_ss_bb}
\end{eqnarray}
\begin{eqnarray}
&-&2\kappa\rho_{aa}+\big[s^*g_0^2 (\rho_{aa}-{\cal{E}}^2 \rho_{00}Y^{-1})VV^{\dag}\nonumber \\
& &- s^*g_0^2V\rho_{bb}V^{\dag} + iE_A^*{\cal{E}}^2\rho_{00}Y^{-1} +~\textrm{h.c.}\big]=0,\hspace{0.2in}
\label{subeq_ss_aa}
\end{eqnarray}
\begin{eqnarray}
2\kappa\rho_{aa}&+&2\gamma \sum_{q}D_q \rho_{bb}D_q^{\dag}\nonumber\\
& &\hspace{0.3in}-\big[ i E_A^* {\cal{E}}^2 \rho_{00}Y^{-1} +\textrm{h.c.}\big]=0,\hspace{0.2in}
\label{subeq_ss_00}
\end{eqnarray}
\end{subequations}
where `h.c.' denotes the Hermitian conjugate. The Eq.\ (\ref{eq:dia_ss}) is still too complicated to solve.
It is noted, however, that except for the second term of Eq.\ (\ref{subeq_ss_00}), all terms are connected only by the dipole transition matrices $V$ and $V^{\dag}$. Therefore, one may expect that Eq.\ (\ref{eq:dia_ss}) can be greatly simplified if the density matrices can be simultaneously diagonalized with both $V$ and $V^{\dag}$.
Note that both $V$ and $V^{\dag}$ are not square matrices. The meaning of diagonalizing these non-square matrices is to make only the diagonal components of these matrices have non zero values as specified in Eq.\ (\ref{eq:VVd_dia}) below \cite{ref:MS_Transform}.

\section{Stationary-State Solution of Master Equation in Natural Basis}\label{sec:sec3}

The system of a degenerate two-level atom interacting with an arbitrary-polarized {\em classical} field can be transformed to a set of independent non-degenerate two-level systems by an unitary transformation, known as the Morris-Shore transformation \cite{ref:MS_Transform}. As shown in Ref.\ \cite{ref:Yudin_pra}, $VV^{\dag}$ and $V^{\dag}V$ become diagonalized simultaneously under this transformation and can be written as
\begin{eqnarray}
VV^{\dag}&=&\sum_{i=1}^{2F+1}\lambda_{i,(a)}^{2}|(a)i\rangle\langle(a)i|,\nonumber\\
V^{\dag}V&=&\sum_{j=1}^{2F+3}\lambda_{j,(b)}^{2} |(b)j\rangle\langle(b)j|.
\label{eq:VVd_nb}
\end{eqnarray}
The states $|(a)i\rangle$ and $|(b)j\rangle$ are the eigenvectors of $VV^{\dag}$ and $V^{\dag}V$, respectively, with $\lambda_{i,(a)}^2$ and $\lambda_{j,(b)}^2$ corresponding eigenvalues, respectively. It can be proven that the eigenvalues are real non-negative numbers and that the sets $\{\lambda_{i,(a)}^2\}$ and $\{\lambda_{j,(b)}^2\}$ of non-zero eigenvalues  coincide with each other.

\begin{figure}
\includegraphics[width=3.4in]{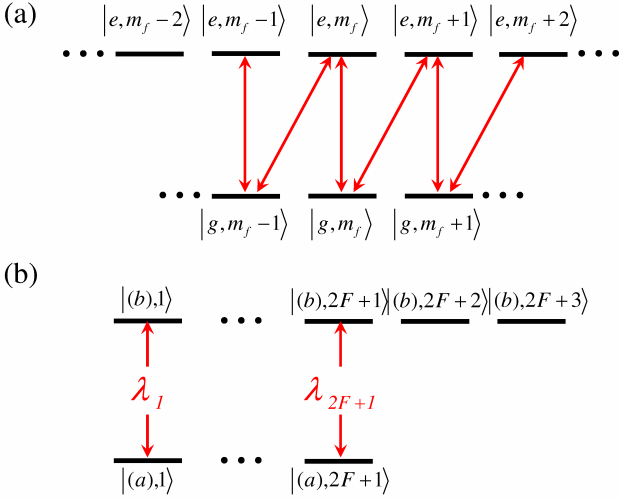}
\caption{(a) Conventional Zeeman basis and transition linkage by polarization $\hat{\textbf{e}}$. (b) Natural basis and the corresponding transition strength $\lambda_i$. Two sublevels of the upper state manifold are uncoupled.}
\label{fig:fig2}
\end{figure}

Accordingly, as shown in Ref.\ \cite{ref:Yudin_pra}, we can write the dipole interaction operators $V$ and $V^\dagger$ in a diagonal form in this new basis, called the natural basis \cite{ref:Yudin_pra}, as
\begin{equation}\label{eq:VVd_dia}
V=\sum_{i} \lambda_i |(a)i\rangle\langle(b)i|, ~~V^{\dag}=\sum_{i} \lambda_i |(b)i\rangle\langle (a)i|.
\end{equation}
where $\lambda_{i,(a)}=\lambda_{i,(b)}=\lambda_i$ for $i=1,2,\ldots,2F+1$ and $\lambda_{j,(b)}=0$ for $j=2F+2$ and $2F+3$, which implies that no transition is allowed for $|(b)2F+2\rangle$ and $|(b)2F+3\rangle$ sublevels [see Fig.\ref{fig:fig2}(b)].

In the conventional Zeeman basis, all degenerate sublevels are linked together by dipole transitions [see Fig.\ref{fig:fig2}(a)].  In the natural basis, however, one ground-state sublevel $|(a)i\rangle$ is coupled to only one excited-state sublevel $|(b)i\rangle$ with a transition strength $\lambda_i$ [see Fig.\ref{fig:fig2}(b)], like a transition by linear polarization in the Zeeman basis. Indeed, for linear or circular polarization, the Zeeman basis coincides with the natural basis.

It is shown in Ref.\ \cite{ref:Yudin_pra} that we can uniquely determine two Hermitian matrices $\cal{A}$ and $\cal{B}$ which satisfy
\begin{eqnarray}
&{\cal{A}}V=V{\cal{B}},~~{\cal{A}}=\sum_q D_q {\cal{B}}D_q^{\dag}
\label{eq:def_AB} \\
&[{\cal{A}}, ~VV^{\dag}]=0,~~[{\cal{B}}, ~V^{\dag}V]=0.
\label{eq:comm_AB}
\end{eqnarray}
The stationary-state solution of the density matrix can then be expressed with these matrices. From Eqs.\ (\ref{eq:VVd_dia})-(\ref{eq:comm_AB}) we can show that these two matrices are also diagonal in the natural basis with the same set of non-zero eigenvalues just like $V$ and $V^{\dag}$. Therefore, in the natural basis these matrices are expressed as
\begin{equation}\label{eq:dia_AB}
{\cal{B}}=\sum_{j=1}^{2F+3} \nu_j |(b)j\rangle\langle(b)j|
,~~{\cal{A}}=\sum_{i=1}^{2F+1} \nu_{i} |(a)i\rangle\langle(a)i|,
\end{equation}
where $\nu_i$'s are eigenvalues of $\cal{A}$ for $i=1,2,\ldots,2F+1$ and $\nu_{2F+2}=\nu_{2F+3}=0$.

From Eqs.\ (\ref{eq:el_pol}), (\ref{Dq-defined}) and (\ref{V-defined}), it can be seen that matrices $V$ and $V^\dagger$ and thus $\cal{A}$ and $\cal{B}$ depend only on the polarization (elliptic angle $\epsilon$) and the atomic level structure (angular momentum $F$). They do not depend on the
other parameters such as detunings, intensity, coupling constant, etc. Moreover, although we are dealing with $F\leftrightarrow F+1$ transition, the matrices $\cal{A}$ and $\cal{B}$ can be found for arbitrary transitions. The details of finding these two matrices are well documented in Ref.\ \cite{ref:Yudin_pra} and references therein, and therefore, in this paper we only give the explicit expressions for matrix elements of $\cal{A}$ and $\cal{B}$ in Appendix \ref{append_mat_elem}.

Substituting Eqs.\ (\ref{eq:def_AB}) and (\ref{eq:comm_AB}) into Eq.\ (\ref{eq:dia_ss}) and after some straightforward algebra with an assumption that $\rho_{bb}$ is proportional to $\cal{B}$, we finally obtain the diagonal elements of the density matrix in closed forms as
\begin{subequations}
\label{eq:fin_dia_ss}
\begin{equation}
\rho_{bb}=\eta {\cal{E}}^2 g_0^2 {\cal{B}}
+O({\cal{E}}^4)\label{subeq_fin_bb}
\end{equation}
\begin{equation}
\rho_{aa}=\eta {\cal{E}}^2 |E_A|^2
(VV^{\dag})^{-1}{\cal{A}}+O({\cal{E}}^4)
\label{subeq_fin_aa}
\end{equation}
\begin{equation}
\rho_{00}=\eta
({Y}{Y}^{\dag})(VV^{\dag})^{-1}{\cal{A}}+O({\cal{E}}^2),
\label{subeq_fin_00}
\end{equation}
\end{subequations}
where $\eta$ is a normalization constant. The results show that $\rho_{bb}$ is indeed proportional to $\cal{B}$, which is self-consistent with the above assumption, and that all the sub-density matrices $\rho_{ij}$ are diagonal in the natural basis.  Furthermore, it is possible to prove that this solution is unique in the same way as described in Ref.\ \cite{ref:Yudin_1st}.

As mentioned previously, the dominant order of $\cal{E}$ is ${\cal{E}}^0$ for $\rho_{00}$, and ${\cal{E}}^2$ for $\rho_{bb}$ and $\rho_{aa}$. The normalization constant $\eta$, determined by the condition $\textrm{Tr}[\rho]=1$, is given by
\begin{equation}
\eta=\frac{1}{|E_A E_C|^2 \alpha_0 -2\textrm{Re}[E_A E_C]\alpha_1
g_0^2 +\alpha_2 g_0^4}+O({\cal{E}}^2),
\end{equation}
with coefficients
\begin{eqnarray}
&\alpha_0=\textrm{Tr}[(VV^{\dag})^{-1}{\cal{A}}],~~\alpha_2
=\textrm{Tr}[VV^{\dag}{\cal{A}}],\nonumber\\
&\alpha_1=\textrm{Tr}[{\cal{A}}]=\textrm{Tr}[{\cal{B}}].
\label{eq:def_alpha}
\end{eqnarray}
The analytic expressions for these coefficients are given in Appendix \ref{append_a012}.

\section{Excitation Spectra and Equivalence to a Non-degenerate Two-Level System}\label{sec:sec4}

In the previous section, we have derived the analytic expression for the stationary density matrix for the system of a degenerate two-level atom coupled to an elliptical-polarized quantized cavity field. We can now obtain the information on the emission from the system from this stationary density matrix. There exist two output channels in our system as shown in Fig.\ \ref{fig:fig1}: The cavity transmission ($T_{\rm cav}$) is characterized by $\kappa$ while the atomic spontaneous emission ($T_{\rm sp}$) is characterized by $\gamma$. These two types of emission can be described by using the results from Eqs.\ (\ref{eq:fin_dia_ss})-(\ref{eq:def_alpha}) as
\begin{eqnarray}\label{eq:T_cavity}
 T_{\rm cav}&=& \textrm{Tr}[a^{\dag}a\rho]=\textrm{Tr}[\rho_{aa}]\nonumber \\
& =& \frac{{\cal{E}}^2 |E_A|^2 }{|E_A E_C|^2 -2\textrm{Re}[E_A
E_C] (\alpha_1 / \alpha_0)g_0^2 +(\alpha_2 /\alpha_0)
g_0^4}\nonumber \\ \nonumber\\
&&  +O({\cal{E}}^4),
\end{eqnarray}
\begin{eqnarray}\label{eq:T_spon}
T_{\rm sp}&=&\textrm{Tr}[\,\sum_{q} D_q^{\dag}D_q
\rho]=\textrm{Tr}[\rho_{bb}] \nonumber \\
&=& \frac{{\cal{E}}^2 (\alpha_1 /\alpha_0 )g_0^2 } {|E_A
E_C|^2-2\textrm{Re}[E_A E_C](\alpha_1 /\alpha_0)g_0^2 + (\alpha_2
/\alpha_0)g_0^4} \nonumber \\ \nonumber\\
&&+O({\cal{E}}^4).
\end{eqnarray}
They are the expressions for the excitation spectra for the cavity and the atomic emission channels, respectively, {\em i.e.}, each describes the normalized emitted power from the corresponding output channel as a function of the driving laser frequency.

When the field is circularly polarized ($\epsilon=\pm\pi/4$), the two spectra coincide with those of the non-degenerate two-level case \cite{ref:Carmichael_ss_sol,ref:Brecha_ss_sol}, by using Eq.\ (\ref{apeq_lim_ratio}).
\begin{equation}
T_{\rm cav}(\epsilon =\pm \frac{\pi}{4})=
{\cal{E}}^2\Big|\frac{E_A}{E_A E_C -g_0^2}\Big|^2 +O({\cal{E}}^4),
\end{equation}
\begin{equation}
T_{\rm sp}(\epsilon =\pm \frac{\pi}{4})=
{\cal{E}}^2\Big|\frac{g_0}{E_A E_C -g_0^2}\Big|^2 +O({\cal{E}}^4).
\end{equation}
Moreover, the structures of Eqs.\ (\ref{eq:T_cavity}) and (\ref{eq:T_spon}) look similar to those of the non-degenerate two-level case. By letting\begin{equation}
g'=\sqrt{\frac{\alpha_1}{\alpha_0}}\,g_0,
\end{equation}
the denominators of Eqs.\ (\ref{eq:T_cavity}) and (\ref{eq:T_spon}) can be written as
\begin{equation}\label{eq:approx_two_1}
|E_A E_C - g'^{2}|^2 +\delta(\epsilon) g'^4,
\end{equation}
where $\delta(\epsilon)\equiv {\alpha_0\alpha_2}/{\alpha_1^2 -1}$, which is much smaller than unity as to be shown below.
Only when the last term of the above expression is negligible, the excitation spectra for two output channels become of the same form as those of the non-degenerate two-level system with $g'$ interpreted as an effective coupling constant of the system.

We can find the precise condition under which we can neglect the last term of Eq.\ (\ref{eq:approx_two_1}). The first term of Eq.\ (\ref{eq:approx_two_1}) can be expanded by using the definition of $E_{A,C}$ as
\begin{eqnarray}
|E_A E_C -g'^2 |^2 &=& (\Delta_C \Delta_A -g'^2)^2 +(\Delta_A \kappa)^2 +(\Delta_C  \gamma)^2~~~ \nonumber\\
& &~+\big[(\kappa \gamma +g'^2)^2 -g'^4 \big].
\end{eqnarray}
The lower bound of the above expression is
\begin{eqnarray}
|E_A E_C -g'^2 |^2\geq4g'^2 \kappa \gamma,~~~~~~~~ &{\rm if}& \kappa \gamma \leq g'^2, \\
|E_A E_C -g'^2 |^2\geq(g'^2 +\kappa\gamma )^2,~~ &{\rm if}& \kappa \gamma \geq g'^2.
\end{eqnarray}
Both are combined to a single condition regardless of the value of $\kappa \gamma/g'^2$ as
\begin{equation}
|E_A E_C -g'^2 |^2\geq4g'^2 \kappa \gamma
\end{equation}
from which we can deduce that the condition for neglecting the last term of Eq.(\ref{eq:approx_two_1}) is
\begin{equation}
\frac{4\kappa \gamma }{g'^2} \geq\frac{4\kappa\gamma}{g_0^2} \gg
\delta(\epsilon). \label{eq:approx_cond}
\end{equation}
Here we use the inequality $g_0 \geq g'$ from Eq.\ (\ref{apeq_a012_inequality}).

\section{Discussion}

\subsection{Accuracy of Two-Level Approximation}
\begin{figure}
\includegraphics[width=3.4in]{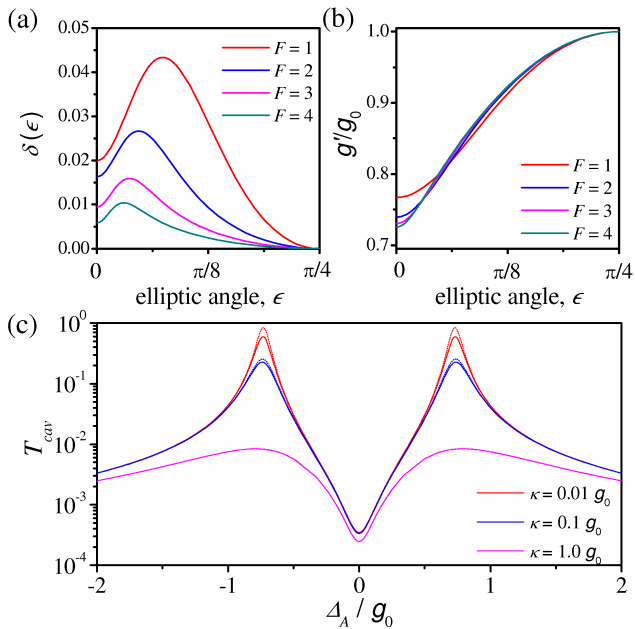}
\caption{(Color Online) (a) Plot of $\delta(\epsilon)$ for various $F$. (b) Variation of $g'/g_0$ as elliptic angle $\epsilon$. (c) Excitation spectrum of the cavity transmission for linear polarization ($\epsilon=0$) when $F=3$. For a fixed value of $\gamma/g_0=0.1$, the value of $\kappa/g_0$ is varied. Solid line represents the spectrum given by Eq.\ ({\ref{eq:T_cavity}}) while the dotted line shows the equivalent non-degenerate two-level system with a coupling constant equal to $g'$.}
\label{fig:fig3}
\end{figure}

In Table \ref{tab:table1}, we present the maximum value of $\delta(\epsilon)$ for various $F$ with relevant atomic species. The variation of $\delta(\epsilon)$ as a function of  $\epsilon$ is shown in Fig.\ \ref{fig:fig3}(a) for some $F$ values. For circular polarization, $\delta(\epsilon)$ becomes zero, consistent with the fact that the system becomes a non-degenerate two-level system in the steady state due to optical pumping. We also present the effective coupling constant $g'$ as a function of elliptic angle $\epsilon$ in Fig.\ \ref{fig:fig3}(b). As mentioned before, when $\epsilon=\pm\pi/4$, $g'$ coincide with $g_0$, and $g'$ has its minimum value when $\epsilon=0$.

\begin{table}
\caption{\label{tab:table1} The maximum values of $\delta(\epsilon)$ for various angular momenta $F$ and corresponding atomic species.}
\begin{ruledtabular}
\begin{tabular}{ccc}
$F$&$\delta_{\rm max}
$ & relevant atomic species\footnote{$\textrm{D}_2$ line of these atoms have $F\leftrightarrow (F+1)$ closed transition.}\\
\hline
1 & 0.043 & $^1 \textrm{H}$\\
2 & 0.027 & $^7 \textrm{Li}$, $^{23} \textrm{Na}$, $^{39}
\textrm{K}$,
$^{41} \textrm{K}$, $^{87} \textrm{Rb}$\\
3 & 0.016 & $^{85} \textrm{Rb}$\\
4 & 0.010  & $^{133} \textrm{Cs}$ \\
\end{tabular}
\end{ruledtabular}
\end{table}

Since $\delta$ is of order of 0.01, the condition in Eq.\ (\ref{eq:approx_cond}) can be safely satisfied if $4\kappa \gamma/g_0^2$ is order of 0.1 or larger. This requirement can be met in a weak coupling regime as well as in intermediate and strong coupling regimes. Some example are given below. If the coupling is too strong, however, the condition is not satisfied.

In Fig.\ \ref{fig:fig3}(c), the excitation spectrum of the cavity transmission, given by Eq.\ (\ref{eq:T_cavity}), of the degenerate two-level system when driven by linear polarization ($\epsilon=0$) is compared with that of the non-degenerate two-level system whose coupling constant is the same as $g'$ of the degenerate two-level system. Both exhibit a double-peak structure, also known as the normal mode splitting. For the same ${\cal{E}}=1$ and $\gamma/g_0=0.1$, the two spectra are compared for three different values of $\kappa$ when $F=3$, for which $\delta(0)\sim 0.01$. When $\kappa/g_0=0.01$ ({\em i.e.}, too strong coupling), the condition in Eq.\ (\ref{eq:approx_cond}) is not satisfied, and consequently, we see a noticeable discrepancy between the two spectra (red solid line vs.\ red dotted line). For other values of $\kappa$ in Fig.\ \ref{fig:fig3}(c), however, the condition is well satisfied, and the two spectra also agree well with each other.

\subsection{Physical Interpretation of Effective Coupling Constant $g'$ and Small parameter $\delta$}

The analytic expressions for $g'$ and $\delta(\epsilon)$ can be derived explicitly by using Eqs.\ (\ref{apeq_a0})-(\ref{apeq_a2}). However, we gain more physical insights by expressing $g'$ and $\delta(\epsilon)$ in the natural basis. With substitution of Eqs.\ (\ref{eq:VVd_nb}) and (\ref{eq:dia_AB}) in Eq.\ (\ref{subeq_fin_aa}), we can express $\rho_{aa}$ in terms of the populations  in the natural basis. We define $\pi_i^{(a)}$ to be the ratio of the population in $|(a)i\rangle$ sublevel to the total ground-state population $\rho_{aa}$:
\begin{equation}
\pi_i^{(a)} \equiv \frac{\nu_i /\lambda_i^2}{\sum_i \nu_i
/\lambda_i^2},~~\sum_i \pi_i^{(a)}=1,
\end{equation}
From Eqs.\ (\ref{eq:VVd_nb}), (\ref{eq:dia_AB}) and (\ref{eq:def_alpha}), we have
\begin{equation}
\alpha_0=\sum_i \nu_i/\lambda_i^2,~\alpha_1=\sum_i\nu_i,~\alpha_2=\sum_i\nu_i\lambda_i^2.
\end{equation}
Then, the $g'$, and $\delta(\epsilon)$ can be rewritten as,
\begin{equation}
g'=\sqrt{\frac{\alpha_1}{\alpha_0}}g_0=\sqrt{\sum_i{\lambda_i^2 \pi_i^{(a)}}}g_0=\sqrt{\overline{\lambda^2}}g_0,
\end{equation}
\begin{eqnarray}
\delta(\epsilon)&=&\alpha_0\alpha_2/\alpha_1^2-1 \nonumber\\
&=&\frac{\Big(\sum_i \lambda_i^4 \pi_i^{(a)}\Big)-\Big(\sum_i \lambda_i^2 \pi_i^{(a)}\Big)^2}{\Big(\sum_i \lambda_i^2 \pi_i^{(a)}\Big)^2}\nonumber\\
&=&\left[\frac{(\Delta \lambda^2)}{\overline{\lambda^2}}\right]^2,
\end{eqnarray}
where $\overline{x}\equiv\sum_i \pi_i^{(a)}x_i$ the population-weighted average of $x$ in the natural basis and $(\Delta x)^2 \equiv \overline{x^2}-\overline{x}^2$ the variance of $x$.
Therefore, the effective coupling constant $g'$ is scaled to $g_0$ by the root-mean-square of the transition strength $\lambda_i$ while the mean is given by the population-weighted average in the natural basis. Likewise, $\delta(\epsilon)$ is the ratio of the variance to the square of the population-weighted average of transition {\em probability} $\lambda_i^2$. In the statistics terminology, $\delta(\epsilon)$ is the same as the square of the coefficient of variation for the transition probability. With this interpretation of $\delta(\epsilon)$, we can have an alternative understanding why $\delta(\pm\pi/4)=0$ for circular polarization: it is due to the vanishing dispersion $\Delta\lambda^2$ since only one $\pi_i^{(a)}$ is nonzero in this case.

\begin{figure}
\includegraphics[width=3.4in]{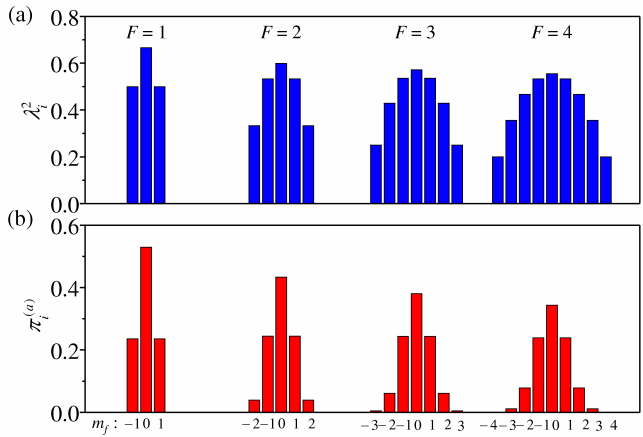}
\caption{Plot of $\lambda_i^2$ and $\pi_i^{(a)}$ for various $F $ values when $\epsilon=0$. As $F$ increases, the difference of $\lambda_i^2$ between three most-contributing sublevels ($i=-1,0,1$) decreases, and accordingly, the dispersion $\Delta\lambda^2$ decreases.}
\label{fig:fig4}
\end{figure}

As shown in Figs.\ \ref{fig:fig3}(a) and \ref{fig:fig3}(b), $\delta(\epsilon)$ decreases as the angular momentum $F$ increases while the change of $g'$ is negligible. This means that the variance of the transition probability decreases as the $F$ increases whereas the average of it does not change much. This can be understood by considering the case of linear polarization, in which the dependence on $F$ is pronounced. When $\epsilon=0$, the transition probability $\lambda_i^2$ becomes,
\begin{equation}
\lambda_i^2 =\frac{(2F+2-i)i}{(F+1)(2F+1)},
\end{equation}
with $i=1, 2, \ldots , 2F+1$. It has a maximum value $(F+1)/(2F+1)$ for $i=F+1$, which corresponds to $m_F=0$  in the  Zeeman basis. The difference between the maximum value and the second maximum value is $1-1/(F+1)^2$, which converges to unity as $F$ increases. Likewise, the difference between the maximum value of $\pi_i^{(a)}$ and the second maximum value of $\pi_i^{(a)}$ also converges to unity as $F$ increases. Therefore, the variance of the transition probability decreases as $F$ increases. Fig.\ \ref{fig:fig4} shows the plot of $\lambda_i^2$, and $\pi_i^{(a)}$ as $F$ is varied. As $F$ increases, the difference of $\lambda_i^2$ between three most-contributing sublevels ($i=-1,0,1$) decreases.

\subsection{Possible Application to Cavity-QED Experiments}
We have shown that the steady state properties of a degenerate two-level Jaynes-Cummings system in the weak excitation limit is approximately the same as that of a simple non-degenerate two-level system with an effective coupling constant. This result can be applied to many cavity QED experiments which should consider linear or elliptic polarization. For example, if one employs special cavities such as microspheres \cite{microsphere-cavity-QED}, microtoroids \cite{microtoroid-cavity-QED} and photonic crystal microcavities \cite{PBG-cavity-QED}, the optical pumping to a cycling transition cannot be fulfilled because circular polarization is not supported there due to the special geometry of supported modes. Therefore, it is necessary to consider the multi-sublevels of atom and coherence among them, making the analysis very complicated. One can still use linear polarization with these cavities and by using our result the original system can be transformed to a system of a non-degenerate two-level atom with a single effective coupling constant.

Another important point to note is that the effective coupling constant can be a new parameter which can be continuously controlled. So far, there has been no way to control the coupling constant continuously in the cavity QED experiments with a single atom. With our result, however, the effective coupling constant can be continuously changed simply by varying the elliptic angle of a probe laser. In particular, with appropriate choice of $g_0$ and $\kappa$, it should be possible to control the effective coupling constants from the weak coupling to the strong coupling regime, and thus one can study the quasi-eigenstates of the atom-cavity composite near the transition point from the weak to strong coupling regime. Such studies are of considerable interest related to an exceptional point \cite{EP} and its associated branch point singularity in the parameter space of the system described by a non-Hermitian Hamiltonian \cite{non-Hermitian}.

\section{CONCLUSION}\label{sec:sec5}
We have obtained an analytic solution for the stationary density matrix of an open Jaynes-Cummings system composed of a degenerate two-level atom and an arbitrary-polarized cavity field. The stationary density matrix is obtained up to the first-order of driving-field intensity. We employ a natural basis formed by the eigenstates of the dipole interaction matrices and show that the stationary density matrix is diagonal in the natural basis. We have also calculated the excitation spectra for the cavity transmission and the atomic spontaneous emission, respectively. Unless the atom-cavity coupling is too strong,  the precise meaning of which we specify, the excitation spectra are shown to be the same as those of an equivalent non-degenerate two-level system with an effective coupling constant $g'$. We show that the effective coupling constant is a function of elliptic angle and angular momentum only, not depending on the other parameters.

The present work is the first to report a full-quantum mechanical analytic solution for the cavity QED system of an degenerate two-level atom with arbitrary polarization under weak excitation. Our results provide a useful theoretical background for better understanding this versatile system and a design basis for future quantum optics experiments including quantum information processing and quantum state engineering. Furthermore, the fact that the effective coupling constant can be continuously controlled suggests a new venue in the cavity QED experiments such as the study of an exceptional point and its associated singular properties in the atom-cavity composite described by a non-Hermitian Hamiltonian.

\begin{acknowledgments}
This work was supported by NRL and WCU Grants.
\end{acknowledgments}

\appendix

\section{Expression for Matrix Elements of $\cal{A}$ and $\cal{B}$}\label{append_mat_elem}

The matrix elements of $\cal{A}$ and $\cal{B}$ are determined only by the elliptic angle $\epsilon$ and angular momentum $F$. For $F\leftrightarrow F+1$ transition, the elements of these matrices can be calculated by using Eq.\ (B8) of Ref.\ \cite{ref:Yudin_pra} as
\begin{eqnarray}
{\cal{A}}_{ij}&=&\sum_{k=-F-1}^{\textrm{Min}[i,
j]}\frac{1}{(i-k)!(j-k)!}\nonumber \\
&&\times~
\sqrt{\frac{(2F+1+i-k)!(2F+1+j-k)!}{(2F+1-i+k)!(2F+1-j+k)!}}\nonumber
\\
&&\times~ C_{F\,i ~(\!F\!+\!1\!)\, -\!k}^{(\!2\!F\!+\!1\!)\,
(\!i\!-\!k\!)}C_{F\, j~ (\!F\!+\!1\!)\, -\!k}^{(\!2\!F\!+\!1\!)\,
(\!j\!-\!k\!)} \Big(\frac{\sin\epsilon}{\sqrt{\cos
2\epsilon}}\Big)^{i+j-2k}, \nonumber\\
\end{eqnarray}
\begin{eqnarray}
{\cal{B}}_{ij}&=&\sum_{k=-F}^{\textrm{Min}[i,
j]}\frac{1}{(i-k)!(j-k)!}\nonumber \\
&&\times~
\sqrt{\frac{(2F+1+i-k)!(2F+1+j-k)!}{(2F+1-i+k)!(2F+1-j+k)!}}\nonumber
\\
&&\times~ C_{(\!F\!+\!1\!)\, i~F
\,-\!k}^{(\!2\!F\!+\!1\!)\,(\!i\!-\!k\!)}C_{(\!F\!+\!1\!)\,j~F
\,-\!k}^{(\!2\!F\!+\!1\!)\,
(\!j\!-\!k\!)}\Big(\frac{\sin\epsilon}{\sqrt{\cos
2\epsilon}}\Big)^{i+j-2k}.\nonumber\\
\end{eqnarray}
The dimension of $\cal{A}$ is $(2F+1)\times (2F+1)$ while $\cal{B}$ is $(2F+3)\times(2F+3)$.

\section{Analytic Expressions for Coefficients $\alpha_0$, $\alpha_1$, and $\alpha_2$}\label{append_a012}

The coefficients $\alpha_0$, $\alpha_1$, and $\alpha_2$ in Eq.\ (\ref{eq:def_alpha}) depend on elliptic angle $\epsilon$ and angular momentum $F$ only. For arbitrary polarization $\hat{\mathbf{e}}$, as defined by Eq.\ (\ref{eq:el_pol}), and for the transition $F\leftrightarrow F+1$, these coefficients can be expressed in explicit form by using Eq.\ (79) of Ref.\ \cite{ref:Yudin_pra} as
\begin{equation}
\alpha_0 =\frac{1}{\cos 2\epsilon}\sum_{l=0}^{F} C_{2l}^2 P_{2l}
\left(1/\cos 2\epsilon\right)\label{apeq_a0}
\end{equation}
\begin{equation}
\alpha_1 =P_{2F+1}\left(1/\cos 2\epsilon
\right)\label{apeq_a1}
\end{equation}
\begin{equation}
\alpha_2=\cos 2\epsilon  \sum_{l=2F}^{2F+2} D_l^2 P_l
\left(1/\cos 2\epsilon\right),\label{apeq_a2}
\end{equation}
where the $P_l (x)$ denotes the Legendre polynomials, and the
coefficients $C_l$, and $D_l$ are given by Eq.\ (77) of Ref.\ \cite{ref:Yudin_pra} as
\begin{equation}\label{apeq_CL}
C_l=\sqrt{\frac{(2l+1)(2F-l)!(2F+l+1)!}{(2F+1)(4F+1)!}} \end{equation}
\begin{eqnarray}\label{apeq_DL}
D_l&=&\sqrt{(2F+3)(4F+3)}C_{10~2F+10}^{l0}\nonumber\\
& & \hspace{0.5in} \times \left( \begin{array}{ccc}
\!l\!&\!1\!&\!2F+1\!\\
\!F\!&\!F+1\!&\!F+1\!\end{array} \right),
\end{eqnarray}
where the last factor denotes the 6$j$ symbol. From $1/\cos 2\epsilon\geq1$, we note that $\alpha_0$, $\alpha_1$ and $\alpha_2$ should have positive values. When $\epsilon =\pm\pi/4$, these coefficients diverge to infinity, but the ratios between them converge to unity:
\begin{equation}
\lim _{\epsilon \rightarrow \pm\frac{\pi}{4} }\,\frac{\alpha_1}
{\alpha_0} = \lim _{\epsilon \rightarrow \pm\frac{\pi}{4}
}\,\frac{\alpha_2}{\alpha_0}=1.
\label{apeq_lim_ratio}
\end{equation}
From the recursion relation of Legendre polynomials
\begin{equation}
(2l+1)xP_l(x)=(l+1)P_{l+1}(x)+l P_{l-1}(x),
\end{equation}
the coefficients $\alpha_0$ and $\alpha_2$ in Eqs.\ (\ref{apeq_a0})-(\ref{apeq_a2}) become
\begin{eqnarray}
\alpha_0 =\alpha_1 &+&\frac{2F}{2F+1}P_{2F-1} \left(1/\cos 2\epsilon\right)\nonumber \\
&+&\frac{1}{\cos 2\epsilon}\sum_{l=0}^{F-1}C_{2l}^2 P_{2l} \left(1/\cos 2\epsilon\right),
\end{eqnarray}
\begin{equation}
\alpha_2 =\alpha_1 -\frac{2F \cos 2\epsilon}{4F+1}P_{2F}\left(1/\cos 2\epsilon\right).
\end{equation}
From the above equations, these coefficients satisfy the following inequality
\begin{equation}
\label{apeq_a012_inequality}
\alpha_0 \geq \alpha_1 \geq \alpha_2,
\end{equation}
and the equality holds for $\epsilon =\pm\pi/4$ as shown in Eq.\ (\ref{apeq_lim_ratio}).


\begin{references}
\bibitem{ref:JCM_1st} E.\ T.\ Jaynes, and F.\ W.\ Cummings, Proc.\ IEEE {\bf 51}, 89 (1963).

\bibitem{ref:JCM_review} B.\ W.\ Shore, and P.\ L.\ Knight, J.\ Mod.\ Opt.\ {\bf 40}, 1195 (1993); B.\ W.\ Shore, J.\ Mod.\ Opt.\ {\bf 54}, 2009
(2007).

\bibitem{ref:JCM_three_level_general} H.\ I.\ Yoo, and J.\ H.\ Eberly, Phys.\ Rep.\ {\bf 118}, 239 (1985).

\bibitem{ref:JCM_three_damp} G.\ Adam, J.\ Seke, and O.\ Hittmair, Opt.\ Commun.\ {\bf 73}, 121 (1989).

\bibitem{ref:JCM_ladder_4} B.\ Buck, and C.\ V\. Sukumar, J.\ Phys.\ A {\bf 17}, 877 (1984).

\bibitem{ref:Kimble_multi} K.\ M.\ Birnbaum, A.\ S.\ Parkins, and H.\ J.\ Kimble, Phys.\ Rev.\ A \ {\bf 74}, 063802 (2006).

\bibitem{ref:degen_sphere} D.\ Lenstra, G.\ Kurizki, L.\ D.\ Bakalis, and K.\ Banaszek, Phys.\ Rev.\ A {\bf 54}, 2690 (1996).

\bibitem{ref:appl_map} A.\ S.\ Parkins, P.\ Marte, P.\ Zoller, and H.\ J.\ Kimble, Phys.\ Rev.\ Lett.\ {\bf 71}, 3095 (1993).

\bibitem{ref:appl_pump} A.\ D.\ Boozer, R.\ Miller, T.\ E.\ Northup, A.\ Boca, and H.\ J.\ Kimble, Phys.\ Rev.\ A {\bf 76}, 063401 (2007).

\bibitem{ref:appl_pspod} T.\ Wilk, S.\ C.\ Webster, H.\ P.\ Specht, G.\ Rempe, and A.\ Kuhn, Phys.\ Rev.\ Lett.\ {\bf 98}, 063601 (2007).

\bibitem{ref:appl_remp} T.\ Wilk, S.\ C.\ Webster, A.\ Kuhn, and G.\ Rempe, Science {\bf 317}, 488 (2007).

\bibitem{ref:MS_Transform} J.\ R.\ Morris, and B.\ W.\ Shore, Phys.\ Rev.\ A {\bf 27}, 906 (1983).

\bibitem{ref:GOBE} C. Cohen-Tannoudji, in {\em Frontiers in Laser Spectroscopy}, edited by R. Ballan, S. Haroche, and S. Liberman (North-Holland, Amsterdam, 1977).

\bibitem{ref:subDoppler} J. Dalibard, and C. Cohen-Tannoudji, J. Opt. Soc. Am. B {\bf 6} 2023 (1989).

\bibitem{ref:milner_EDS} V.\ Milner and Y.\ Prior, Phys.\ Rev.\ Lett.\ {\bf 80}, 940 (1998).

\bibitem{ref:Yudin_1st} A.\ V.\ Ta\u{\i}chenachev, A.\ M.\ Tuma\u{\i}kin, and V.\ I.\ Yudin, JETP {\bf 83}, 949 (1996).

\bibitem{ref:milner_pra} V.\ Milner, B.\ M.\ Chernobrod, and Y.\ Prior, Phys.\ Rev.\ A {\bf 60}, 1293 (1999).

\bibitem{ref:Yudin_nc} A.\ M.\ Tuma\u{\i}kin and V.\ I.\ Yudin, Sov.\ Phys.\ JETP {\bf 71}, 43(1990).

\bibitem{ref:Carmichael_ss_sol} H.\ J.\ Carmichael, R.\ J.\ Brecha, and P.\ R.\ Rice, Opt.\ Commun.\ {\bf 82}, 73 (1991).

\bibitem{ref:Brecha_ss_sol} R.\ J.\ Brecha, P.\ R.\ Rice, and M.\ Xiao, Phys.\ Rev.\ A {\bf 59}, 2392 (1999).

\bibitem{ref:Yudin_pra} A.\ V.\ Ta\u{\i}chenachev, A.\ M.\ Tuma\u{\i}kin, V.\ I.\ Yudin, and G.\ Nienhuis, Phys.\ Rev.\ A {\bf 69}, 033410 (2004).

\bibitem{microsphere-cavity-QED} D.\ W.\ Vernooy, A.\ Furusawa, N.\ Ph.\ Georgiades, V.\ S.\ Ilchenko, and H.\ J.\ Kimble, Phys.\ Rev.\ A {\bf 57} R2293 (1998).

\bibitem{microtoroid-cavity-QED}  S.\ M.\ Spillane, T.\ J.\ Kippenberg, K.\ J.\ Vahala, K.\ W.\ Goh, E.\ Wilcut, and H.\ J.\ Kimble, Phys.\ Rev.\ A {\bf 71} 013817 (2005).

\bibitem{PBG-cavity-QED} J.\ Vu\u{c}kovi\'{c}, M. Lon\u{c}ar, H. Mabuchi, and A. Scherer, Phys.\ Rev.\ E {\bf 65} 016608 (2001).

\bibitem{EP} T. Kato, {\em Perturbation Theory for Linear Operators} (Springer, New York, 1966).

\bibitem{non-Hermitian} C.\ Dembowski, H.\ -D.\ Gr\"{a}f, H.\ L.\ Harney, A.\ Heine, W.\ D.\ Heiss, H.\ Rehfeld, and A.\ Richter Phys.\ Rev.\ Lett.\ {\bf 86} 787 (2001).

\end{references}
\end{document}